\documentclass[12pt,preprint]{aastex}
\newcommand{\Msun}{~M_\odot}

\newcommand{\kms}{\rm ~km~s^{-1}}

\newcommand{\ml}{~\Msun ~\rm yr^{-1}}

\usepackage{xcolor,nicefrac}
\usepackage{morefloats}

\begin{document}

\title{NEAR INFRARED SPECTROSCOPY OF THE TYPE IIn \\      
SN 2010jl: EVIDENCE FOR HIGH VELOCITY EJECTA}

\author{H. Jacob Borish\altaffilmark{1}, Chenliang Huang\altaffilmark{1}, Roger A. Chevalier\altaffilmark{1}, Benjamin M. Breslauer\altaffilmark{1,2}, Aaron M. Kingery\altaffilmark{1,3}, George C. Privon\altaffilmark{1,4}}
\altaffiltext{1}{Department of Astronomy, University of Virginia, P.O. Box 400325, 
Charlottesville, VA 22904-4325; hjborish@virginia.edu} 
\altaffiltext{2}{Google, 1600 Amphitheatre Pkwy, Mountain View, CA 94043}
\altaffiltext{3}{ERC Incorporated/Jacobs ESSSA Group, NASA Marshall Space Flight Center, Huntsville, AL 35812}
\altaffiltext{4}{Departamento de Astronom\'{i}a, Universidad de Concepci\'{o}n, Casilla 160-C,
Concepci\'{o}n, Chile}

\begin{abstract}
 The Type IIn supernova SN 2010jl was relatively nearby and luminous,
allowing detailed studies of the near-infrared (NIR) emission.
We present $1-2.4~\mu$m spectroscopy over the age range of 36 -- 565 days from the
earliest detection of the supernova.
On day 36, the H lines show an unresolved narrow emission component along with a symmetric broad component
that can be modeled as the result of electron scattering by a thermal distribution of electrons.
Over the next hundreds of days, the broad components of the H lines shift to the blue by $700\kms$, as is also observed in optical lines.
The narrow lines do not show a shift, indicating they originate in a different region.
\ion{He}{1} $\lambda$10830 and $\lambda$20587 lines both show an asymmetric broad emission component, with a shoulder on the blue side that varies in prominence and velocity from $-5500\kms$ on day 108 to $-4000\kms$ on day 219.
This component may be associated with the higher velocity flow indicated by X-ray
observations of the supernova.
The absence of the feature in the H lines suggests that this is from a He rich ejecta flow.
The \ion{He}{1} $\lambda$10830 feature has a narrow P Cygni line, with absorption extending to
$\sim 100\kms$ and strengthening over the first 200 days, and an emission component which weakens with time.
At day 403, the continuum emission becomes dominated by a blackbody spectrum with a temperature of $\sim 1900$ K, suggestive of dust emission.
\end{abstract}


\section{INTRODUCTION}

Type IIn supernovae (SNe IIn) are characterized by narrow optical emission lines and a blue
continuum \citep{schlegel90}.
The presence of a narrow line component at early times is taken to indicate a
dense surrounding circumstellar medium that is heated and ionized by radiation
from the explosion and the continuing interaction.
The presence of wings on the narrow lines, as observed for the Type IIn SN 1998S \citep{leonard00},
can at least in some cases be explained by electron scattering in a mildly optically thick medium \citep{chugai01}.
The  optical luminosity of a Type IIn event can also be explained by shock
interaction with a dense circumstellar medium \citep{chugai94}.
The velocity of the circumstellar mass loss can be deduced from the velocities indicated
by P Cygni line profiles observed in the H Balmer lines; the typical line velocities
are $100-1400\kms$ \citep{kiewe12}.
The mass loss rates implied by the interaction luminosities are $\sim 0.01-0.1\ml$.
High mass loss rates were also derived from infrared emission attributed to circumstellar dust \citep{fox11}.
Roughly $6\%-11\%$ of core collapse supernovae are of Type IIn \citep{smith2011a}. 

These rates of mass loss are much higher than those found in normal stars, and
the place of Type IIn events in stellar evolution is unclear.
The progenitor stars have been linked to luminous blue variables (LBVs)
because these have the requisite mass loss density and outflow velocity during their
eruptive phases \citep[e.g.,][]{smith10,fox11, kiewe12}.
In addition, the progenitor of SN 2005gl is consistent with an LBV
\citep{galyam09}.
A problem is that stars in this evolutionary phase are not expected to undergo
supernova explosions \citep[although see the rotating models of][]{groh13}.
Suggestions for the proximity in time of mass loss and explosion include
wave-driven mass loss \citep{quataert12} and binary-driven mass loss \citep{chevalier12},
but these studies are very speculative.
\cite{soker13} also attributes the dense circumstellar medium around SNe IIn to binary interaction.

The bright Type IIn supernova SN 2010jl
was discovered on UT 2010 November 3.5 by \citet{NP10}.  
It was subsequently confirmed at an unfiltered magnitude of 12.9.  
Pre-discovery images showed that the supernova was present on 2010 October 9.6 \citep{stoll11},
which we use as our zero point in time.
The supernova is located at R.A. $=9^h42^m53\fs33$, Dec. $=+9^\circ29\arcmin41\farcs 8$ (equinox 2000.0).  It lies 2\farcs4 east and 7\farcs7 north of the center of the host galaxy, UGC 5189A. 
We take the distance to the supernova to be 49 Mpc \citep{smith11b} and the reddening to
be $E(B-V)=0.058$ \citep{fransson13}.
\cite{smith11b} used pre-explosion HST imaging to identify a bright, blue point source coincident with the position of the supernova, which they took to imply that the progenitor of SN 2010jl had a mass above 30 $M_{\Sun}$. 
\cite{andrews11} observed the supernova in the {\it Spitzer} 3.6 and 4.5 $\mu$m bands
and the $JHK$ bands
during days $90-108$, finding evidence for 7500 K and 750 K components.
However, \cite{fransson13} have examined the IR data for this age and find the temperature of
the cool component to be $\sim 1800$ K;
this temperature characterizes the cool component for 100's of days.
\cite{fransson13} presented one $0.3-2.4~\mu$m spectrum of SN 2010jl (on day 460),
as well as extensive optical and ultraviolet observations and near infrared (NIR) photometry.
{\it Spitzer} observations were also presented by \cite{fox13}, who find evidence for  circumstellar
emission and a high mass loss rate.
\cite{maeda13} presented a spectrum (optical through NIR) on day 563, finding evidence
for dust formation in the supernova ejecta.

SN 2010jl has also been detected as an X-ray source \citep{immler10,chandra12,ofek14}.
Observations on 2010 December $7-8$ (day 58 since discovery of the supernova)
 showed a hot ($kT>8$ keV) thermal spectrum with an absorbing column $\sim 10^{24}$ cm$^{-2}$, assuming
incomplete ionization of the circumstellar gas and a metallicity 0.3 of solar  \citep{chandra12}.
This column density corresponds to an electron scattering optical depth $\sim 1$.
By 2011 October $17-18$ (day 372), the absorbing column declined by a factor of 3,
but the X-ray emission remained steady and hot ($kT>12$ keV).
However, \cite{ofek14} re-examined the {\it Chandra} data and found them difficult to model.
Hard X-ray observations with {\it NuSTAR} on 2012 October 6 yielded a temperature of
$\sim 18_{-4}^{+6}$ keV \citep{ofek14}, which corresponds to a shock velocity of
$\sim 4000\kms$.

Here we present a set of 10 NIR spectra of SN 2010jl spanning 529 days (Section 2).
The various components contributing to the line and continuous emission are described in
Section 3.
The physical picture implied by the observations is discussed in Section 4 and the
conclusions are in Section 5.

\section{OBSERVATIONS}

Data presented here were obtained using TripleSpec, an $0.9-2.4 ~\mu$m, $R \sim 3000$ spectrograph, on the ARC 3.5 m telescope at Apache Point Observatory \citep{wilson04,herter08}.  The supernova was observed 10 times between 2010 November 15 and 2012 April 26.  Table \ref{obstable} shows the details of each observation.  Day numbers listed in Table \ref{obstable} refer to time elapsed since the first detection of the supernova on 2010 Oct 9.  Integration time refers to the total exposure time for each date.  Individual exposures were 5 minutes in length and nodded in an ABBA pattern along the slit.  We also observed the spectrophotometric standard HD 85377 for the telluric correction and photometric calibration.  The airmass at the time of each observation is given in Table \ref{obstable} for both SN 2010jl and HD 85377.  Sky subtraction was achieved by differencing each pair of A and B exposures.  Sky subtraction, spectral extraction, telluric correction, and photometric calibration were carried out using a modified version of the IDL utility \emph{SpexTool} \citep{cushing04}. 

\begin{figure}[p]
\plotone{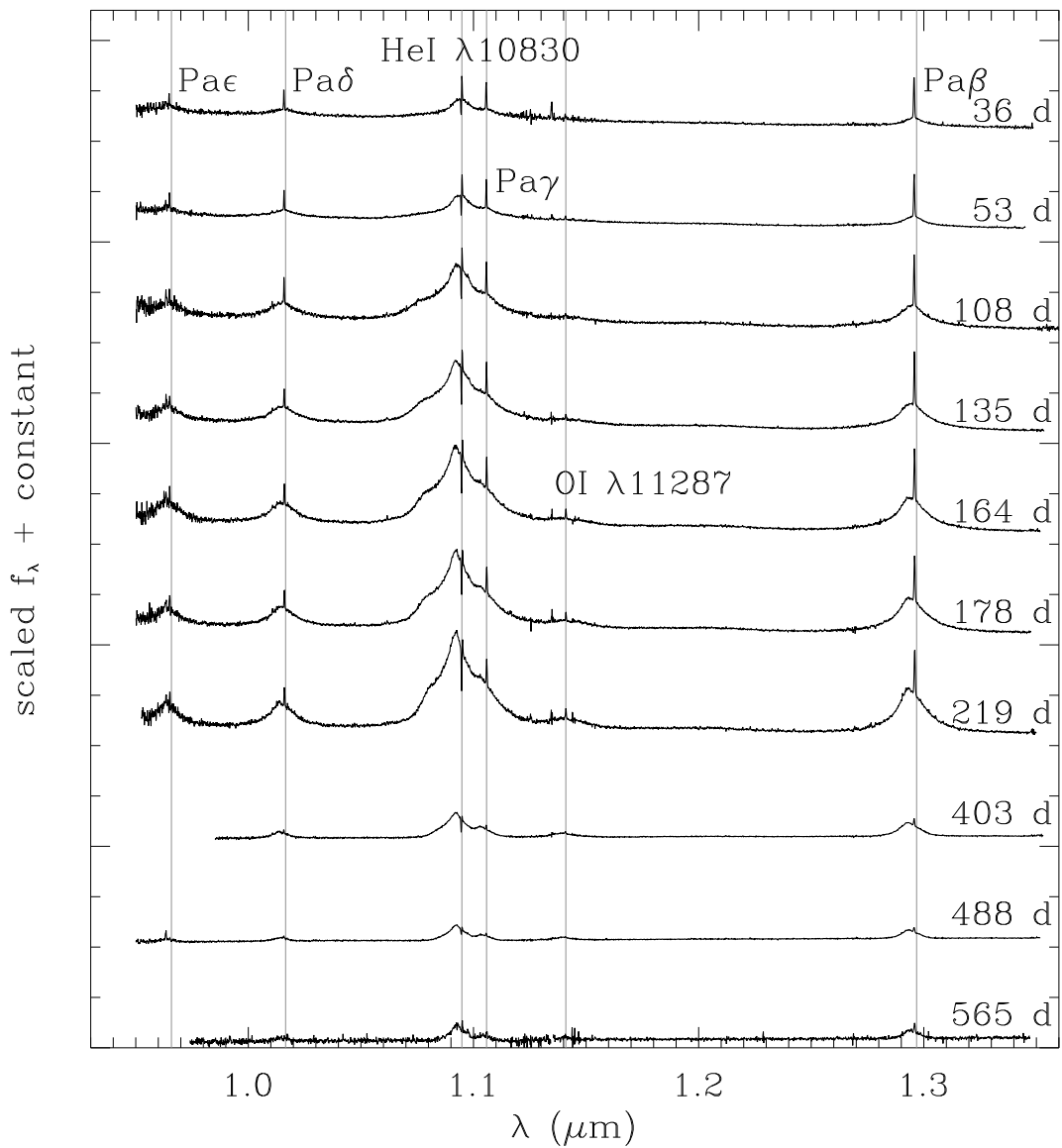}
\caption{$J$ band spectra of SN 2010jl taken at the ARC 3.5m Telescope.  The data have not been corrected for redshift or reddening.  
The spectra have been scaled so that the continuum of each spectrum has the same strength at $1.55~ \mu$m and then displaced vertically for clarity.  The day numbers refer to the earliest observation, 2010 October 9.  Figures \ref{spectrumh} and \ref{spectrumk} were prepared in the same way.}
\label{spectrumj}
\end{figure}

\begin{figure}[p]
\plotone{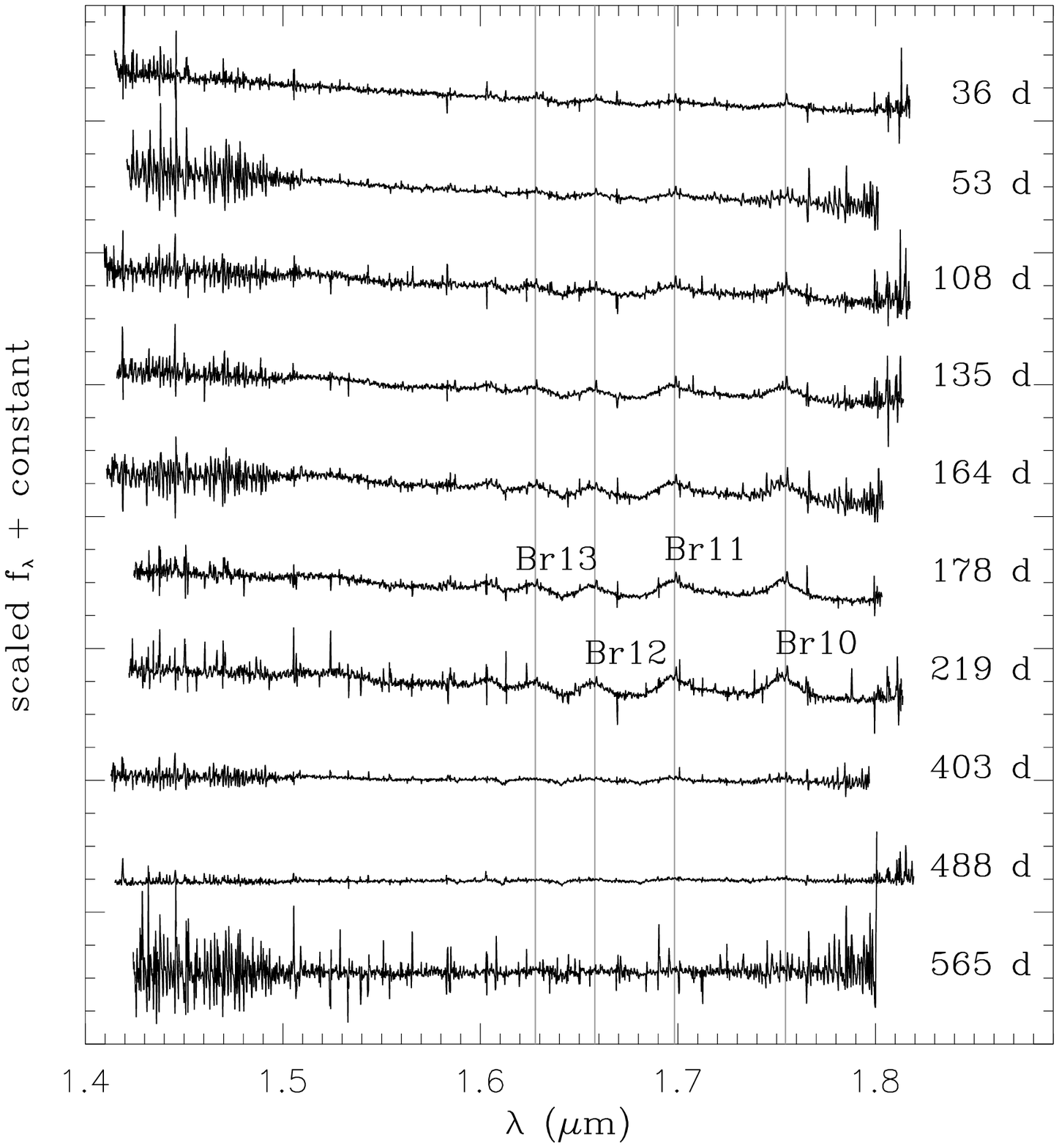}
\caption{$H$ band spectra of SN 2010jl taken at the ARC 3.5m Telescope.}
\label{spectrumh}
\end{figure}

\begin{figure}[p]
\plotone{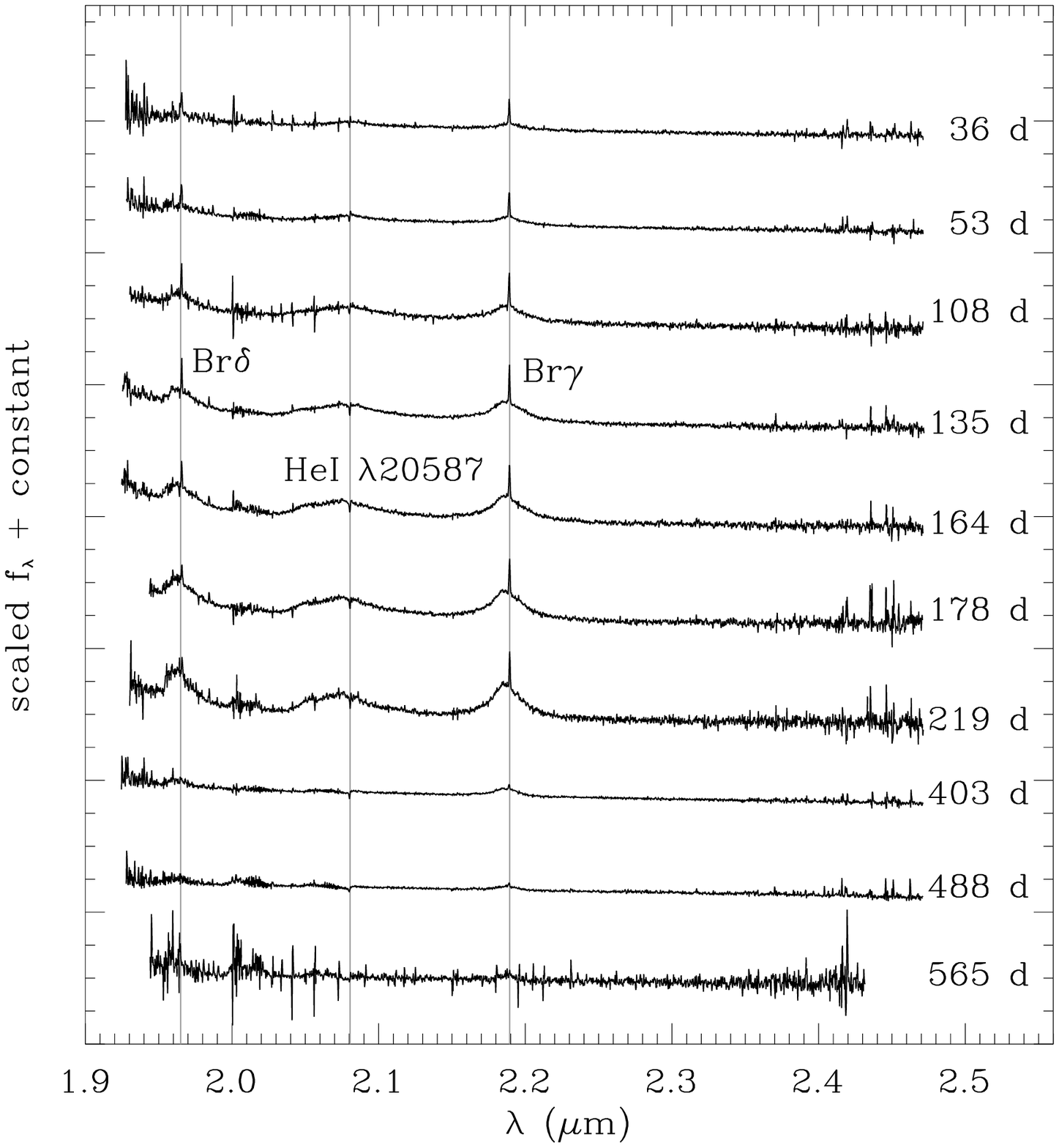}
\caption{$K$ band spectra of SN 2010jl taken at the ARC 3.5m Telescope. }
\label{spectrumk}
\end{figure}

Figures $\ref{spectrumj} - \ref{spectrumk}$  show the spectra in the $J$, $H$, and $K$ bands, respectively.
The spectra are normalized to have the same continuum intensity at 1.55 $\mu$m.
Identifications of the dominant lines are given.

\begin{figure}
\plotone{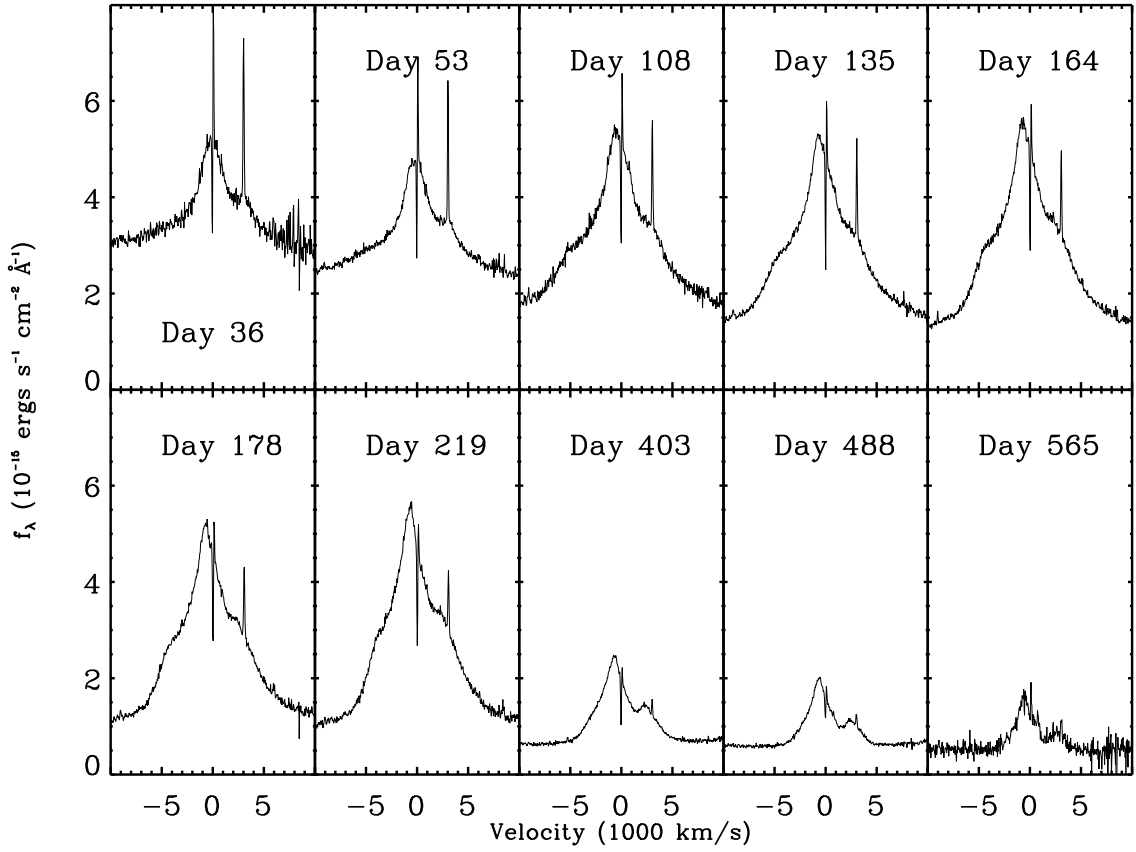}
\caption{Time evolution of the \ion{He}{1} $\lambda$10830 line.  The Pa$\gamma$ line is superposed on the redshifted \ion{He}{1} emission.}
\label{He10evo}
\end{figure}

\begin{figure}
\plotone{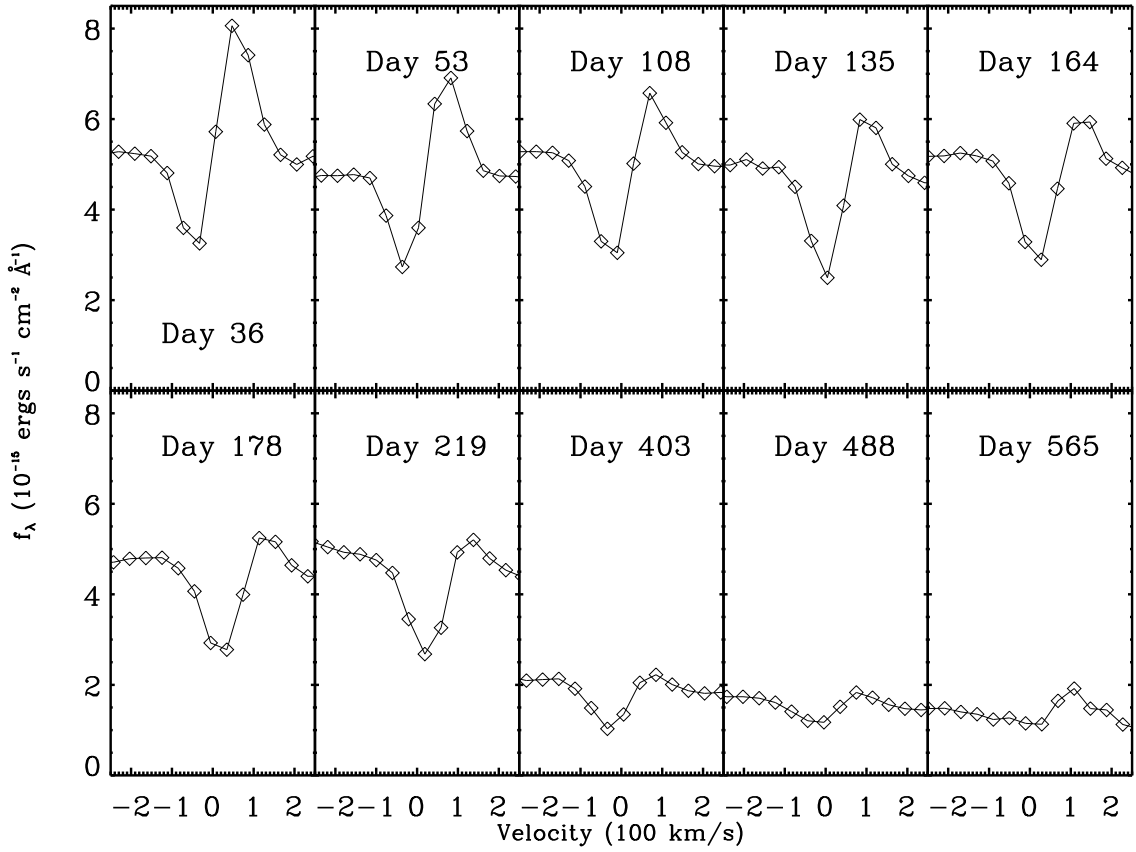}
\caption{Time evolution of the narrow P-Cygni feature in the \ion{He}{1} $\lambda$10830 line.}
\label{pcygevo}
\end{figure}

Variations in seeing and cloud cover are problematic for photometric calibration of any long-slit spectrum. Therefore, the photometric calibrations produced by \emph{SpexTool} are only roughly accurate. Much of the following discussion deals with descriptions of the line or continuum shapes, rendering a precise photometric calibration unimportant. Figures \ref{PBevo}, \ref{He10evo}, \ref{pcygevo}, and \ref{oievo} show the true flux density as revealed by photometric observations collected by \cite{fransson13}. To estimate \emph{H}-band magnitudes at the exact dates of our observations we fit separate $2^{nd}$ degree polynomials to the early time (before 250 days) and late time (after 403 days) \emph{H}-band light curves published by \cite{fransson13}. Because \cite{fransson13} does not publish photometry nearby in time to day 403, we used cubic spline interpolation over the full light curve to estimate the magnitude at 403 days. As a result, the calibration to the photometry for day 403 is somewhat more uncertain than that at other dates. Nevertheless, the behavior of the flux density at day 403 seems to fall in line with the other observations. Once \emph{H} magnitudes had been determined for each TripleSpec observation, we converted each magnitude to a flux density using 
\begin{equation}
H = -2.5 \log(f_{1.63 \mu m}) - 24.860,
\end{equation}
where $1.63 ~\mu$m and $-24.860$ are the effective wavelength and magnitude zero-point of the NIR \emph{H}-band as given by \cite{bessell98}. We then scaled the spectra output by \emph{SpexTool} to the level of the calculated flux density at $1.63 ~\mu$m. 

\section{SPECTRAL COMPONENTS}

\subsection{Continuum Emission}

\begin{figure}[p]
\plotone{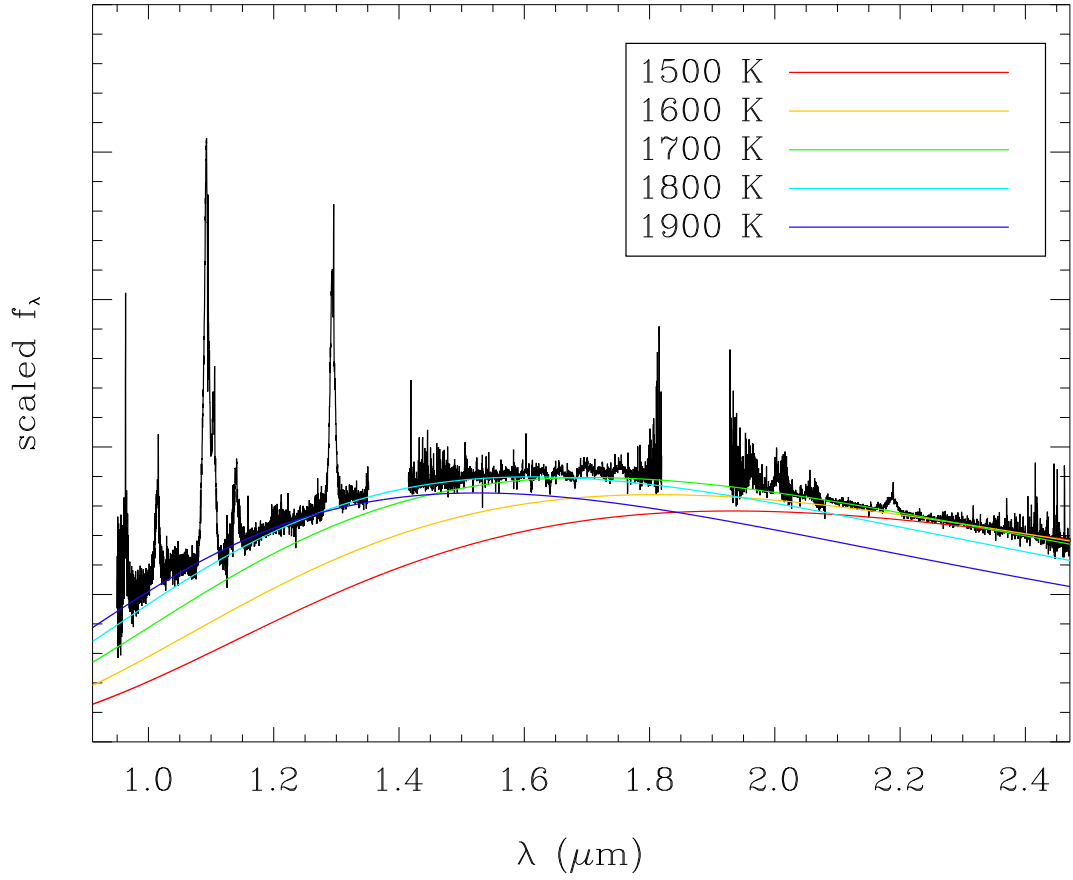}
\caption{The spectrum at 488 days overlaid with blackbody curves. The blackbodies span the range in temperature for which the blackbody peak lies within the near-infrared.}
\label{latebbfit}
\end{figure}

At early times the continuum emission has a slope that is slightly flatter than Rayleigh-Jeans. The flatness of the continuum relative to the blackbody shape indicates the combination of warm and cool components. The reason for this can be seen in the decomposition of the continuum emission by \cite{fransson13}, who model the optical through infrared spectrum and find that a relatively cool ($1800-1900$ K) component is present from early times as well as a warmer ($\sim 7000$ K) component. Because the blackbody peak is in the optical range, temperatures determined from optical observations should be more reliable than those determined from the infrared through day 219. At times in excess of one year, the continuum emission has a local maximum within the wavelength range of our observations. On 2011 November 17 (403 days), the continuum peak is found at about $1.5~\mu$m. By 2012 February 9 (488 days), it has moved to about $1.7~\mu$m. Figure \ref{latebbfit} presents the spectrum at 488 days overlaid with five blackbodies ranging in temperature from 1500 to 1900 K, showing that the observed continuum is flatter than a blackbody. Modifying the blackbody curve to account for the emissivities of various dust grains does not produce a shape which fits the continuum emission because the efficiency of dust emission drops with increasing wavelength. 

Assuming a cooler component contributes to the emission at wavelengths longer than $1.5~\mu$m, we identified selected areas of continuum emission and fit to a blackbody using the IDL fitting package MPFIT \citep{markwardt09,more78} for each observation later than 1 year. The temperatures  of these fits are $\sim 2000$ K and are shown in Table \ref{bbfits}. In Figure \ref{latebbfit}, the best fits to the continuum short of $1.9~\mu$m lie well within $100$ K of $1800$ K. We therefore cite an approximate uncertainty of $\pm 75$ K for these temperatures. This is the uncertainty in the shortest wavelength emission observed in the NIR and does not take into account the evidence for cooler emission at wavelengths longer than $1.9~\mu$m. By fitting the shortest wavelength emission we estimate the temperature of the warmest dust.
Although the shape of the continuum on days 488 and 565 does not fit a pure blackbody, a temperature of $\sim 1900-2000$ K is consistent with emission from hot dust and with the observations of \cite{fransson13}. Such emission has previously been observed in a number of Type IIn supernovae \citep{fox11}. 
\subsection{Hydrogen  Lines}

As can be seen in Figures \ref{spectrumj} -- \ref{spectrumk}, many emission lines of the Paschen and Brackett series of H are detected.
Each line shows 2 distinct components: a narrow component that is close to zero velocity in the host galaxy and a broad component.
Although narrow lines could have their origin in the interstellar medium of the host galaxy, we attribute the emission primarily
to the unshocked circumstellar medium.
One reason is that the fluxes of narrow lines at later times ($t\ga 380$ day) decline to about 1/10 of the fluxes at earlier times,  implying
we can comfortably ignore the contribution of host galaxy emission lines at early times.
Another reason is that interstellar emission lines would also be present in the Balmer series.
Observations of the early H$\alpha$ line show a narrow P Cygni feature \citep{smith11b,fransson13}, indicating that the circumstellar medium
is optically thick in the H Balmer lines. The P Cygni features observed in the H$\alpha$ line indicate an expansion velocity
of $100\kms$ \citep{fransson13}. The Paschen and Brackett series lines do not show absorption, suggesting they are optically thin.
The narrow lines are plausibly formed in a region of  circumstellar mass loss.

\begin{figure}
\plotone{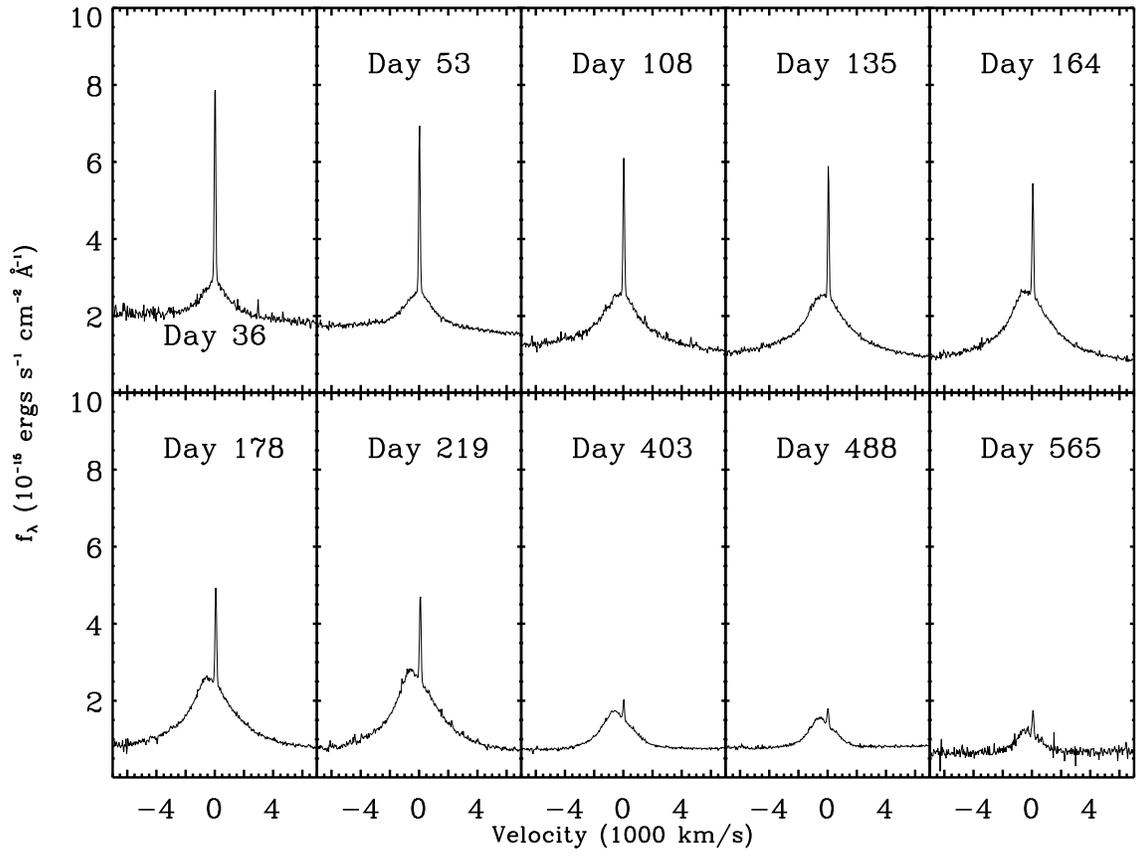}
\caption{Time evolution of the Paschen $\beta$ line.}
\label{PBevo}
\end{figure}

Broadened emission from hydrogen is visible in the data throughout the accessible lines of the Paschen and Brackett series.  The best detected lines are Paschen $\beta$, Brackett $\gamma$, and Paschen $\delta$.  
The shape and evolution of the Paschen $\beta$ line are shown in Figure \ref{PBevo}.  Paschen $\gamma$ is also strong, but is heavily blended with the red side of the broad \ion{He}{1} $\lambda$10830 line.  Paschen $\epsilon$ is sometimes observed, but it lies on the edge of an atmospheric absorption region, making the determination of its shape and strength strongly dependent on changing atmospheric conditions.  Many of the weaker Brackett series lines are identifiable in the data, and corroborate the common line shape of broad H emission. 

At early times, the broad component is fairly symmetric about zero velocity.
In other Type IIn supernovae, a Lorentzian profile gives a reasonable fit to broad Balmer lines \citep{leonard00,smith10}.
Here, the broad Paschen $\beta$ line is moderately well fit by a Lorentzian profile  \citep[see also][for H$\alpha$]{smith11b}.
However, the broad line profile is presumably the result of the thermal particle
velocities and scattering; this situation does not naturally produce a Lorentzian profile.
The broad line profile is naturally produced by electron scattering \citep{chugai01,fransson13}, without the need for Doppler effects due to bulk motion.
The narrow line component is made up of photons that escape without scattering, so its strength relative to the broad component places constraints on the scattering optical depth to the line photon source. 
At early times, the broad component is symmetric, implying that the electron thermal velocities are much
larger than systematic velocities.
At late times, the broad lines are seen to become asymmetric with regard to zero velocity,
although the H$\alpha$ line remains fairly symmetric relative to the line peak \citep{fransson13};
the peak of the broad component shifts to the blue by $\sim 700\kms$, as also occurs
in the optical lines. 
Figure \ref{PBevo} shows that the extended wings in the Paschen $\beta$ line have diminished by day 400.
We take this to indicate that the electron scattering optical depth is decreasing,  which could
occur because the shock wave overruns the scattering region or there is a change in the
ionization of the gas.

\begin{figure}
\plotone{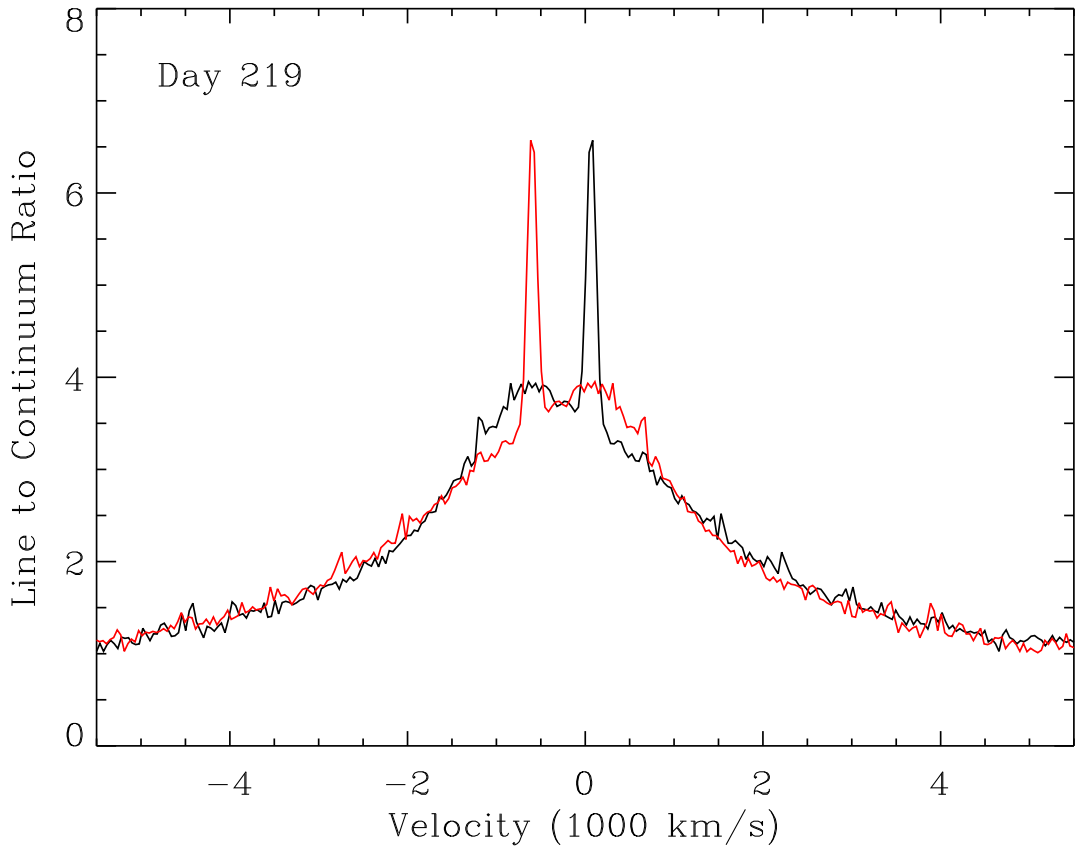}
\caption{Mirroring the profile of Paschen $\beta$ on day 219 about a velocity of $265 \kms$ reveals an asymmetry in the line core when the reflection of the wings is well matched.}
\label{betamirror}
\end{figure}

In addition to finding that the emission lines shift to the blue with time, 
\cite{fransson13} found that the lines remain symmetric about a blueshifted wavelength
during the shift.
They used this property to argue that the line emission and scattering occur in a
comoving slab of material.
In Figure \ref{betamirror} we show the Paschen $\beta$ line on day 219 as well as its reflection about a particular velocity, using
a procedure essentially the same as that used by \cite{fransson13}.
The velocity of the shift is chosen so that the wings of the lines over $1000 - 5000 \kms$ match up.
It can be seen that there is an asymmetry near the peak of the line.
Although the asymmetry is small, it shows that the hypothesis of electron scattering in  a comoving slab is not completely accurate.
\cite{fransson13} presented NIR spectroscopy of the supernova on day 465, finding symmetry of the lines at that time.
Our late time spectra also show that the degree of asymmetry becomes small.

\begin{figure}
\plotone{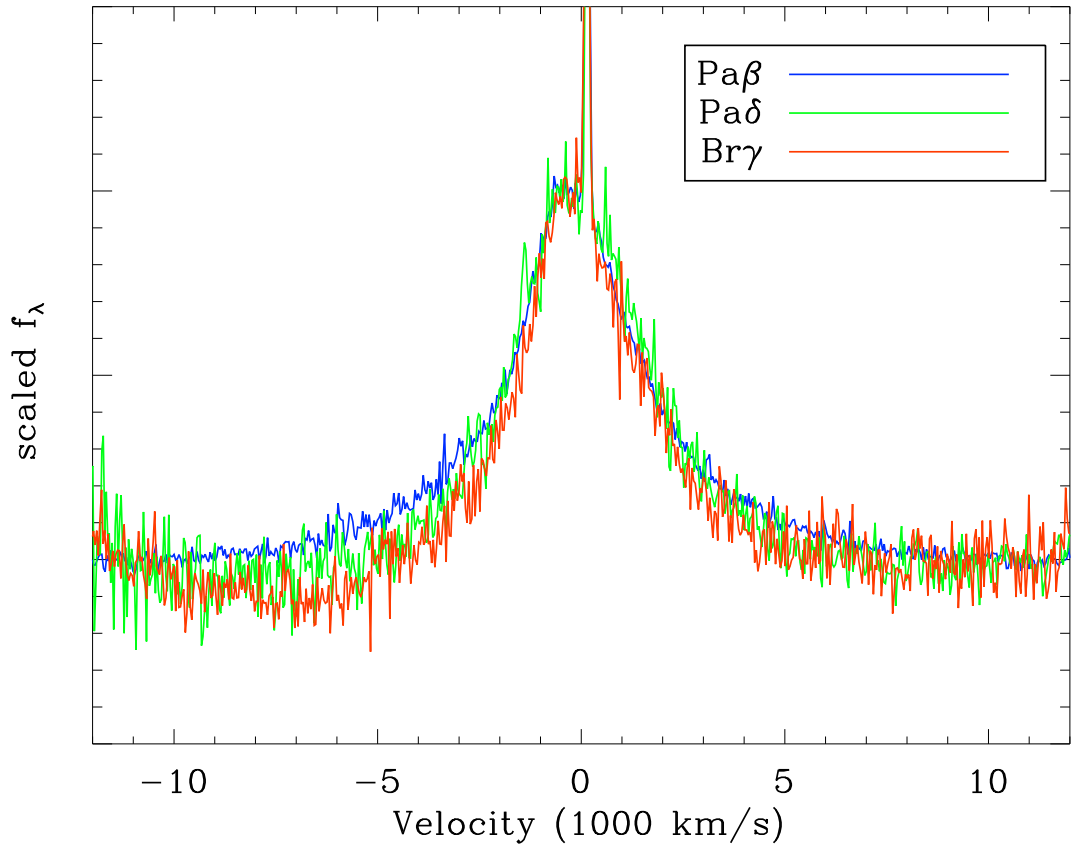}
\caption{Comparison of the different H lines at day 164.  After subtraction of a 4865 K blackbody, a linear background was fit between velocities of $\pm 10,000\kms$ and $\pm 12,000\kms$ and subtracted.  Each line was then scaled to a height of 1.}
\label{Hcompare}
\end{figure}

Figure \ref{Hcompare} compares the broad components of the Paschen $\beta$, Brackett $\gamma$, and Paschen $\delta$ lines
at one time, 164 days.    
It can be seen that the FWHM of the lines are similar.
There are deviations in the line wings, but this aspect is sensitive to the continuum subtraction.
We conclude that there is no strong evidence for differences between the lines.
In the case of SN 1994W, \cite{dessart09} discussed differences between these lines caused by
differences in optical depth.
However, those authors were considering a case where the emission was from shocked, and thus very high density, gas.
In our case, we believe the broad lines are formed in the unshocked circumstellar medium, where the densities are lower.

\begin{figure}
\plotone{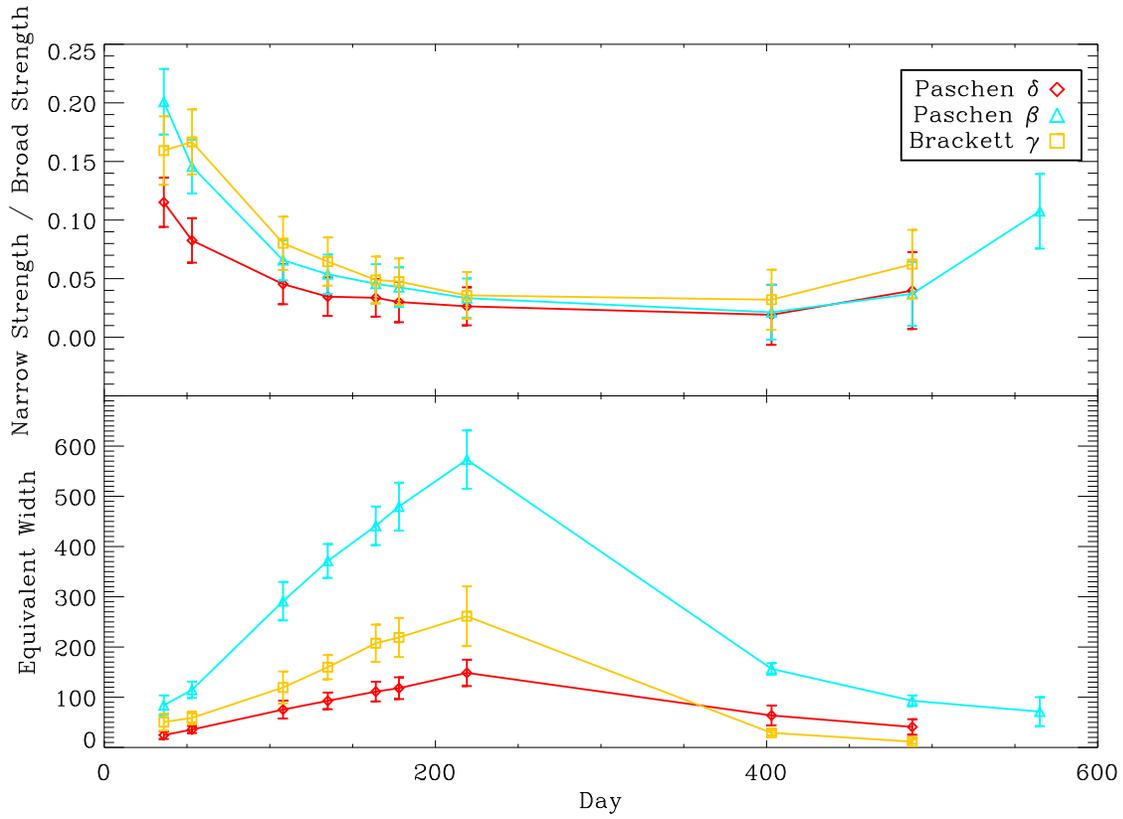}
\caption{Evolution of the equivalent widths and narrow to broad line strengths of several prominent lines. {\bf The equivalent widths plotted include both narrow and broad line flux.}}
\label{HEW1}
\end{figure}

Figure \ref{HEW1} and Table \ref{n2b} shows the evolution of the flux in the narrow component to that in the broad component of Paschen $\beta$, Paschen $\delta$, Brackett $\gamma$, and OI $\lambda$11287.
Figure \ref{HEW1} and Table \ref{EWT} show the evolution of the equivalent widths of the prominent lines.
In order to measure both narrow component strength to broad component strength ratios and equivalent widths, a linear fit to the continuum in the vicinity of each emission line was found.
The ratios of narrow component strength to broad component strength were measured by direct integration over each component after subtraction of the continuum fit.
The quoted uncertainties were found by allowing the full width of the narrow component to vary by one pixel in each direction while also taking into account the noise in the continuum about the line.
It can be seen that the ratio of narrow to broad line flux decreases with time.
Equivalent widths were measured by directly integrating over the ratio of line to continuum.
The quoted uncertainties were found by allowing the constant parameter in the linear continuum fit to vary by $1\sigma$ in both directions.
There is an initial approximately linear increase of the equivalent width with time, followed by a decline between days 220 and 400.
\cite{zhang12} and \cite{fransson13} found a similar evolution for the H$\alpha$ line.

\subsection{He lines}

\begin{figure}
\plotone{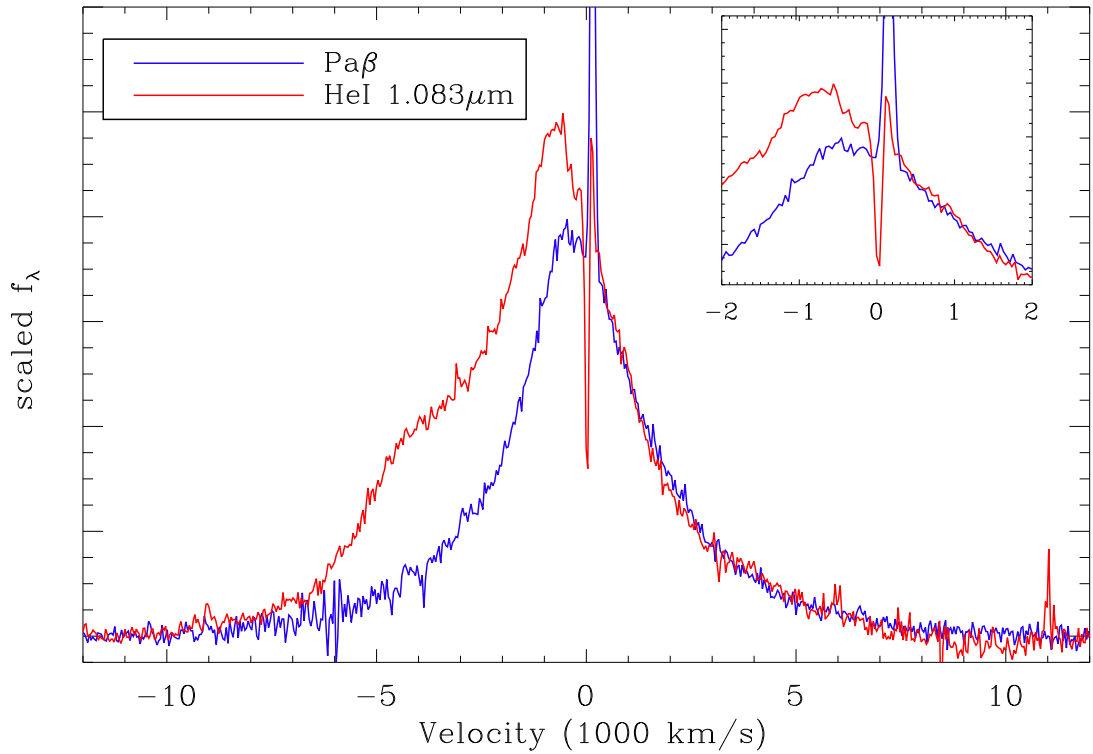}
\caption{Comparison of the He $\lambda$10830 to the H Pa$\beta$ line at day 178.  After subtraction of a 4865 K blackbody, the Paschen $\beta$ profile was scaled to the strength of the Paschen $\gamma$ and subtracted, as described in the text. 
The inset shows the difference in velocity between the peak of the helium profile and that of of the hydrogen profile.}
\label{HHecompare}
\end{figure}

Broad He emission features are detected at $1.0830 \mu$m and $2.0587 \mu$m. The evolution of the He lines, especially \ion{He}{1} $\lambda$10830, shows some similarities to the H lines, although
there are significant differences (Figure \ref{He10evo}).
The broad component shows an increasing equivalent width, in the same way as the H lines (Figure \ref{HEW1}).
A comparison of the \ion{He}{1} $\lambda$10830 line with the Paschen $\beta$ line on day 164 is shown in Figure \ref{HHecompare}.
It can be seen that both lines show extended wings going out $\sim 8000-9000\kms$.
On the red side of the \ion{He}{1} $\lambda$10830 line, there is the superposition of the Paschen $\gamma$ line, which
has both broad and narrow emission components like the Paschen $\beta$ line.
 Assuming the line profile of Paschen $\gamma$ is identical to Paschen $\beta$, in Figure \ref{HHecompare} we use Paschen $\beta$ as a template to obtain a Paschen $\gamma$ subtracted He profile.  The shift and scale of the Paschen $\beta$ template is chosen in order to best fit the template narrow peak to the narrow peak of Paschen $\gamma$.  
The peak of the broad component shows a shift to the blue by $\sim 1000 \kms$ in contrast to the $\sim 700 \kms$ shift in the peak of the H lines  (Figure \ref{HHecompare} inset).
As described below, we attribute this difference to an additional component in the He line.

\begin{figure}
\plotone{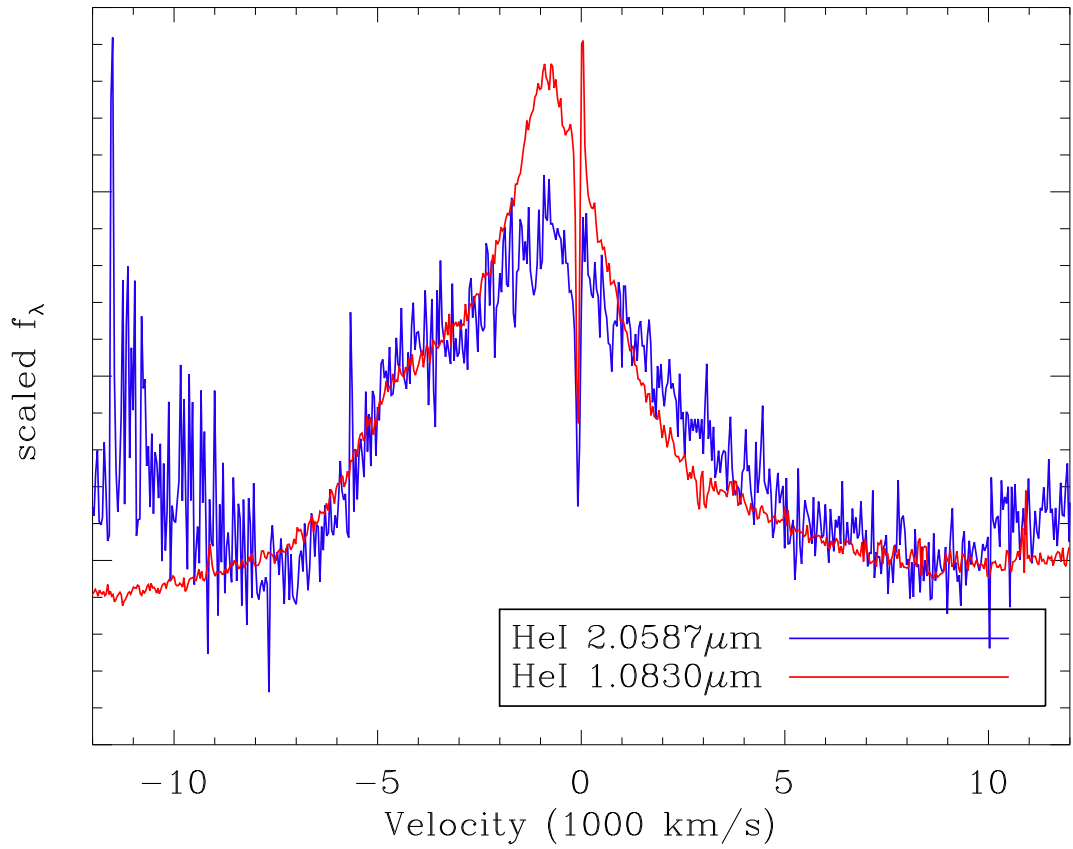}
\caption{Comparison of the \ion{He}{1} $\lambda$10830 line to the $\lambda$20587 line at day 164.  After subtraction of a 4865 K blackbody, a linear background was fitted between velocities of $\pm 7,000\kms$  and $\pm 9,000\kms$  and subtracted. Each line was then scaled to a height of 1.}
\label{Hecompare}
\end{figure}

\begin{figure}  
\plotone{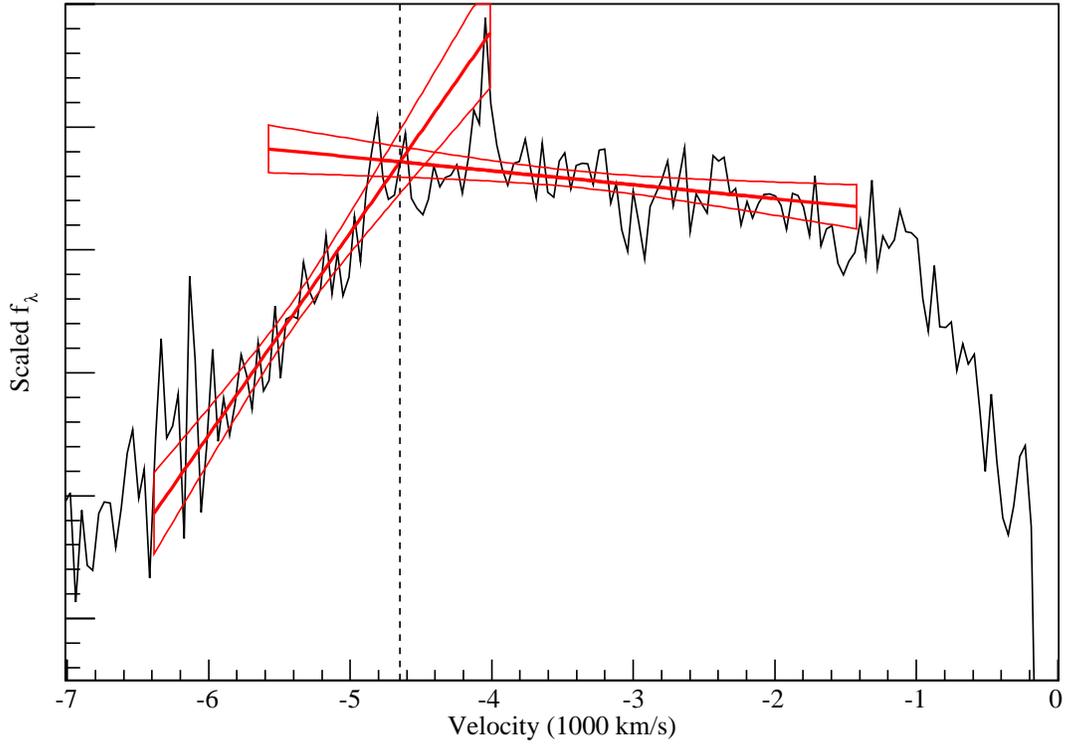}
\caption{The residual of a subtraction of Paschen $\beta$ from He $\lambda$10830  (Figure \ref{HHecompare}),  showing the distribution of blueshifted excess He emission on day 178. We find the velocity of the inflection point in the emission (at $ \sim -5000\kms$) by fitting lines to each region adjacent to the inflection point. Also overplotted are regions of 95\% confidence about the linear fits. The quoted uncertainty in the inflection point is given by the width of the overlap of the uncertainty regions of each fit.}
\label{178_sh}
\end{figure}

\begin{figure}
\plotone{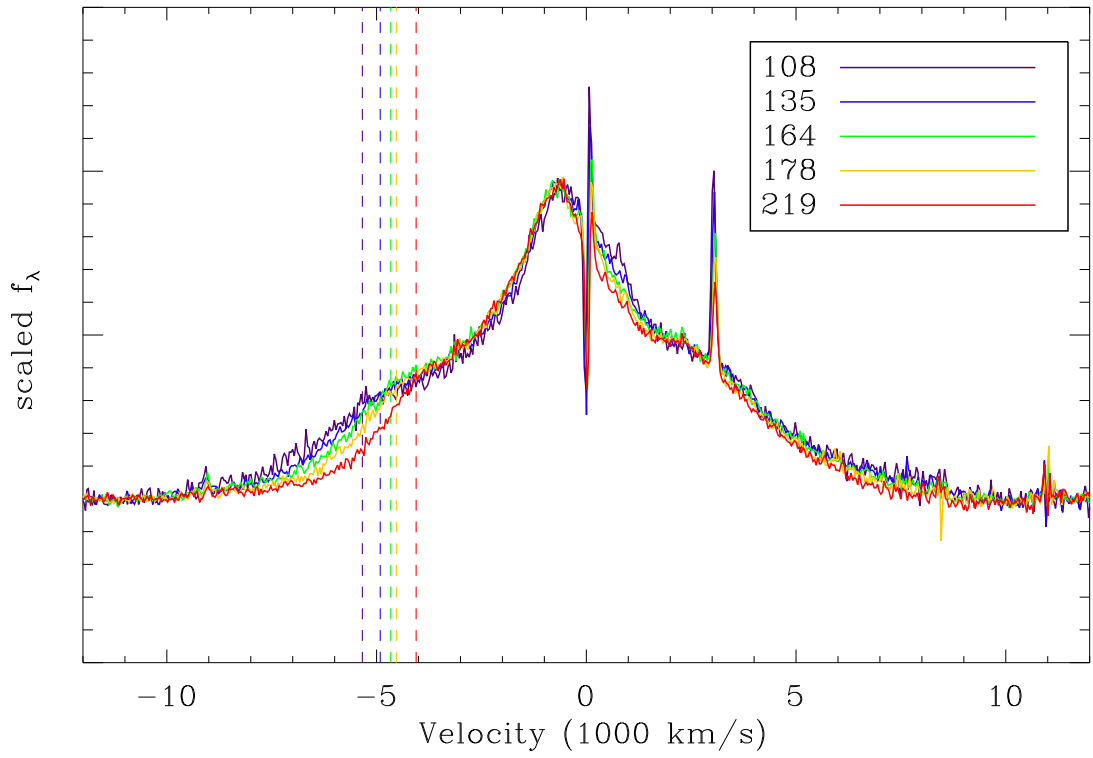}
\caption{Plot of broad \ion{He}{1}  $\lambda$10830 emission from day $108-219$ showing evolution of the broad shoulder. Dashed lines show the velocity of the break in the shoulder at each time.}
\label{shoulderevoln}
\end{figure}

\begin{figure}
\plotone{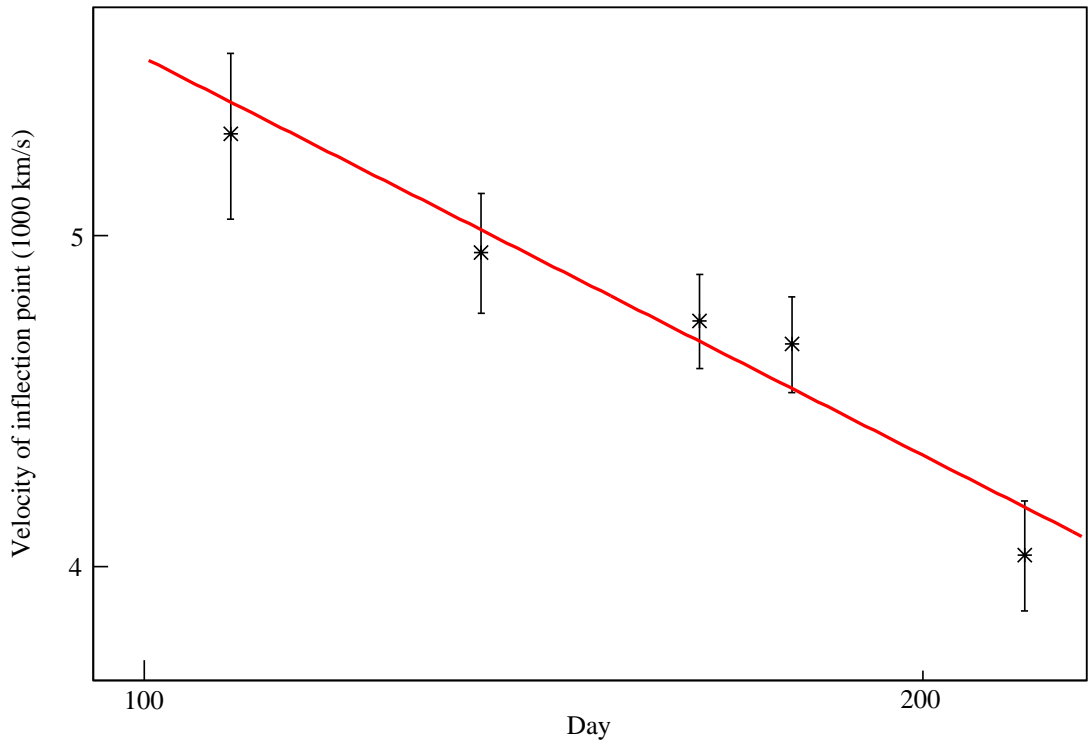}
\caption{Evolution of the velocity of the inflection point  in the broad \ion{He}{1}  $\lambda$10830 emission profile over days $108-219$.  This is a $\log-\log$ plot and the line shows an evolution of velocity $\propto t^{-0.39}$. Error bars have been calculated based on the uncertainty shown in figure \ref{178_sh}.}
\label{velocityevoln}
\end{figure}

 On the blue side on day 164, there is a shoulder feature in the \ion{He}{1} line that is not present in the H line.
In   Figure \ref{HHecompare}, the Paschen $\gamma$-subtracted \ion{He}{1}  line is overplotted
with the Paschen $\beta$ line so that the line profiles match at positive velocities.
We have used a similar procedure to search for excess emission on the other dates of observation.
On days 36 and 53, the feature is not discernable;  on days 402 and 488, there is not a well defined shoulder feature, but there is weak excess He emission that peaks at $\sim -2500\kms$.   
The feature grows and fades in relative strength roughly in the same way as the broad H lines.
One possibility for the origin of this feature is that it is a separate line with only a broad component and no narrow component.
However, a comparison of the \ion{He}{1} $\lambda$10830 line to the \ion{He}{1} $\lambda$20587 line (Figure \ref{Hecompare}) shows that the shoulder feature is present in both lines.  
The broad wings are likely due to electron scattering; components with different temperatures are possible, but
this would not produce a shoulder structure on one side of the line profile.
The implication is that the feature is due to material with systematic velocities up to $\sim 5000\kms$ towards the observer.
Figure \ref{178_sh} shows the excess blueshifted emission in the He I $\lambda$10830 line relative to the Paschen $\beta$ line; the emission extends to at least $-6000\kms$.
In addition, the shoulder feature of excess emission moves to the red with time (Figure \ref{shoulderevoln}),   
which is the opposite of the evolution of the H broad features.
The deceleration of the break in the shoulder feature  follows a velocity evolution $t^{-0.39\pm 0.08}$
(Figure \ref{velocityevoln}). The error bars in the shoulder-break velocities shown in figure \ref{velocityevoln} are estimated by measuring the overlap of the $1\sigma$ uncertainty regions for each linear fit component shown in figure \ref{178_sh}.

In addition to the broad line feature, there is a narrow P Cygni feature in
the \ion{He}{1} $\lambda$10830 line.
Figure \ref{pcygevo}  shows that initially (day 36) the emission component dominates over the narrow absorption.  By day 53, the 2 components are at nearly equal strength and at later times,
the absorption component dominates.
Since the resolution of the observations is about $100 \kms$, we place an upper limit of $100 \kms$  on the wind velocity of the CSM.  The width of the absorption line does not change with the time of our observations.  The same is true for the H$\alpha$ line \citep{fransson13}. 
This is indicative of the formation of the circumstellar medium in an approximately steady
flow. If the material were launched in an eruptive event, the CSM velocity would
increase roughly linearly with radius, and over time the observed emission-line widths should increase.
This is not observed, at least within the spectral resolution.  
The narrow component in the \ion{He}{1} $\lambda$20587 line is not as well defined
as in the $\lambda$10830 line;
it  appears only in absorption.

\subsection{Other Lines}

\begin{figure}
\plotone{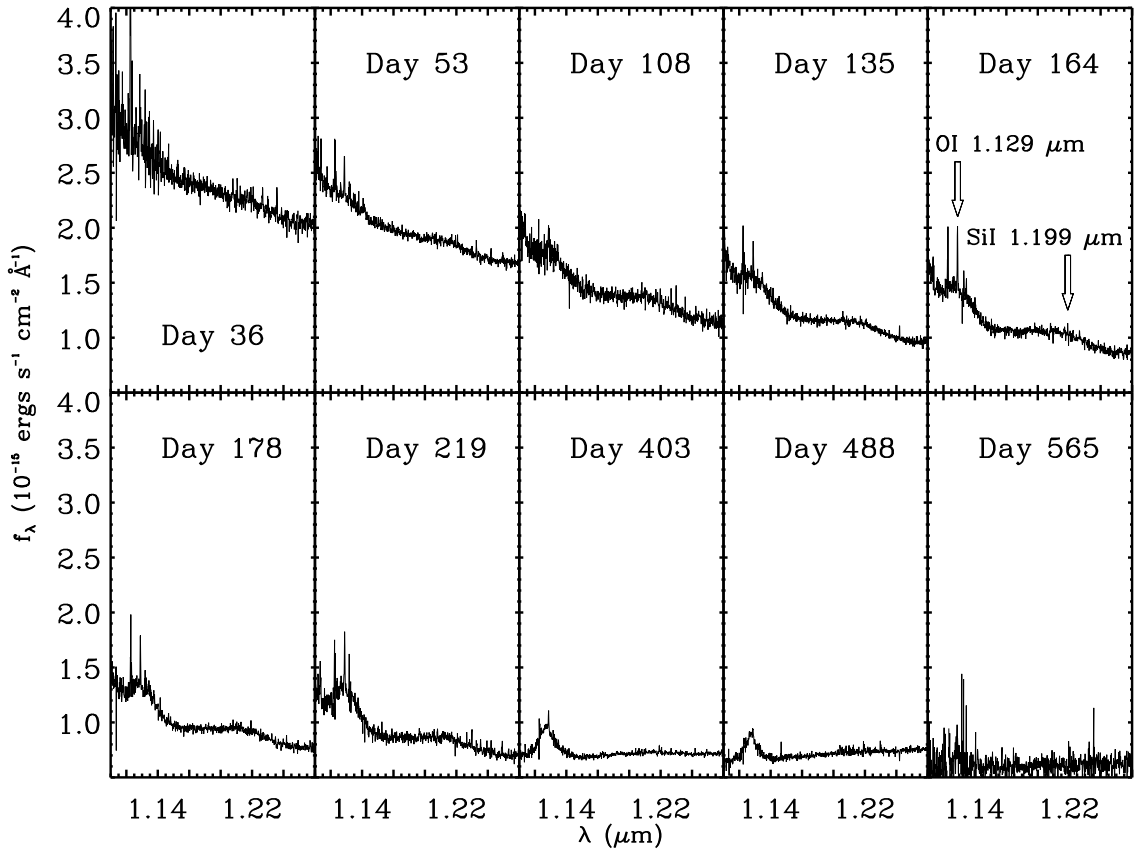}
\caption{Time evolution of the \ion{O}{1} $\lambda$11287 and \ion{Si}{1} $\lambda$11991 features.}
\label{oievo}
\end{figure}

The \ion{O}{1} line at $1.129~\mu$m is in a noisy part of the spectrum where there is confusion with
night sky features.
In addition, it overlaps the red wing of the \ion{He}{1} $\lambda$10830 line, but it
is clearly present (Figure \ref{oievo}).
The line shows a broad  emission component with
properties and an evolution that are comparable to those of the H lines. 
A narrow line may be detected but is uncertain because of the 
contamination by night sky lines.  
The observed emission is compatible with the view that  the \ion{O}{1} line is the result of
pumping by the Ly$\beta$ line \citep{fransson13}.

\begin{figure}
\plotone{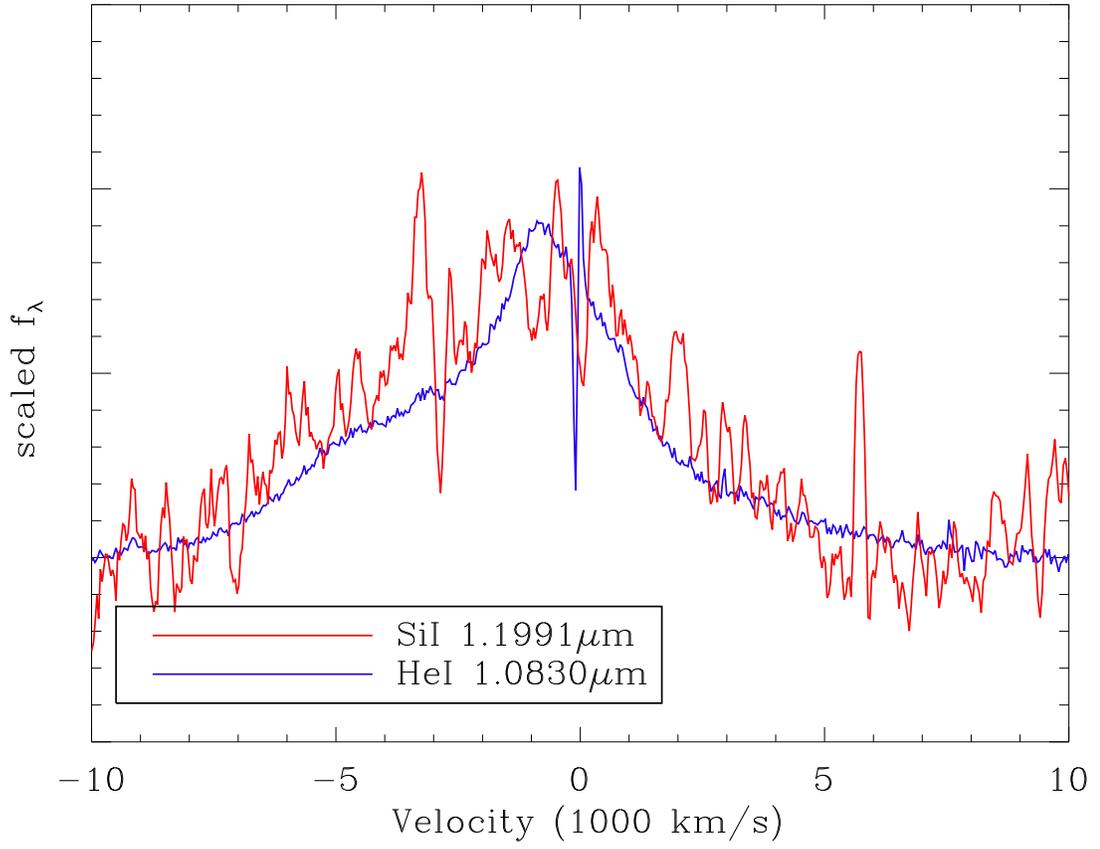}
\caption{\ion{Si}{1} $\lambda$11991 overlaid with \ion{He}{1} $\lambda$10830 on day 135.}
\label{SiI1.19}
\end{figure}

In addition, there is a line present at $\sim 1.2~\mu$m (Figure \ref{oievo}).  
It has a broad, but no narrow, component.
We tentatively identify it as  \ion{Si}{1} $1.1991-1.2270 ~\mu$m.  
We also considered \ion{Mg}{1} $\lambda$11828, but it gave a poor fit to the line profile,
under the assumption that the line profile is comparable to that of \ion{He}{1} $\lambda$10830.
The  line profile, when taken to be the \ion{Si}{1} line, shows that there is emission  associated with
the blue shoulder emission observed in the He lines (Figure \ref{SiI1.19}).

\section{PHYSICAL PICTURE}

Our interpretation of the NIR spectra builds on previous discussions of Type IIn
supernovae \citep{chugai94,leonard00,fransson02,hoffman08,fransson13}.
One component is the
pre-supernova mass loss region with which the supernova is interacting.
The mass loss has an outward
velocity of $100\kms$ and is fairly steady over the region of observation.
The region is photoionized by X-ray radiation coming from the shock interactions.

\begin{figure}
\plotone{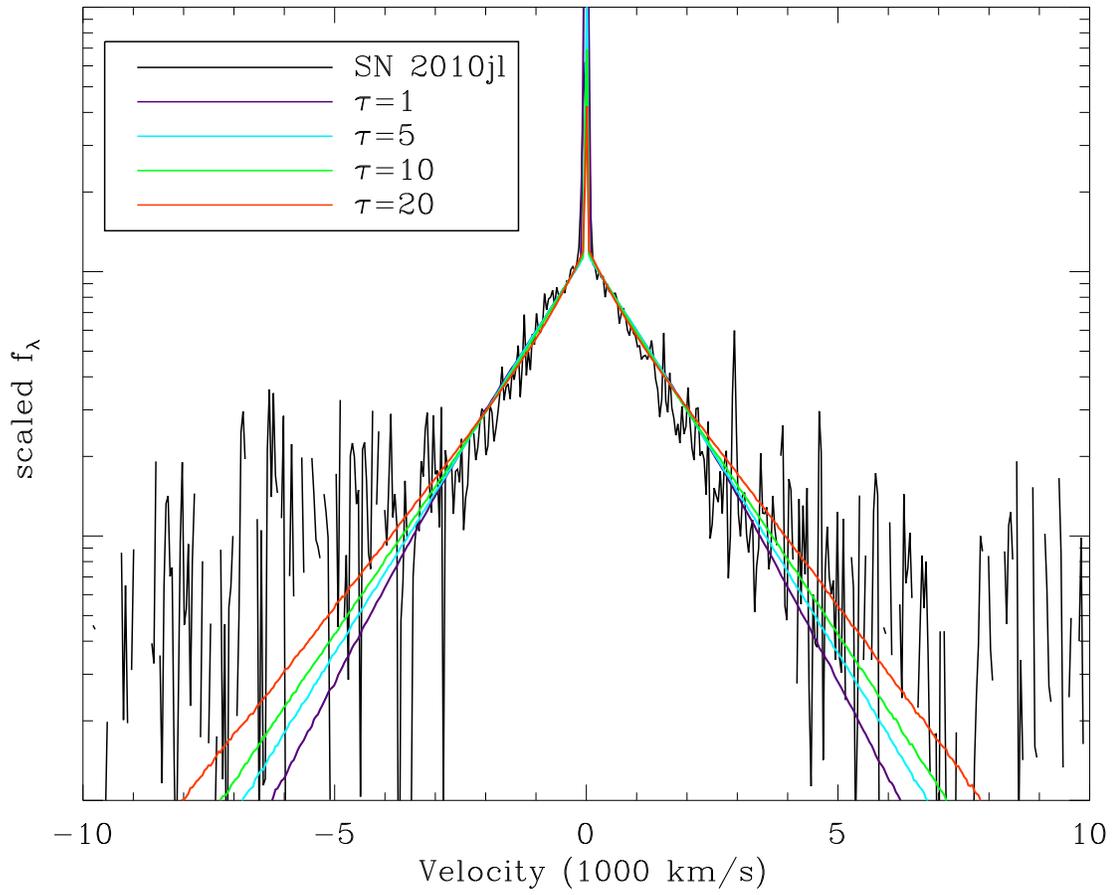}
\caption{Electron scattering profiles overlaid on the broad component of the Paschen $\beta$ line at an age of 36 days. In order to compare the line profile shapes, the model profiles have been scaled to approximately the same FWHM as the Paschen $\beta$ line. }
\label{ESfits}
\end{figure}

In our first epoch of NIR spectra, the H line profiles show a narrow component due
to the slow wind and a broad symmetric component that we attribute to electron scattering
in the slow wind.
The line width reflects a combination of the electron scattering optical depth, $\tau_e$,
and the gas temperature \citep{chugai01,huang14}.
We  initially  assume that the   
emitting gas is the same as the scattering gas, and both are associated with the slow wind
which has a density profile $n\propto r^{-2}$ over an extended region.
The scattering process was modeled by a Monte Carlo code to follow the scattering and
wavelength shifts of line photons.
As in \cite{fransson13}, the broad line component is well-reproduced by the electron scattering
line profile (Figure \ref{ESfits}). 
The model assumed spherical symmetry, with an absorbing sphere for an inner boundary $R_i$.
As shown in \cite{huang14}, the FWHM of the scattered line, which has an approximately
exponential shape, depends on the optical depth to $R_i$, $\tau_e$,  and the temperature
$T$.
Provided that the ratio of outer radius to $R_i$ is $>10$, the results  are not sensitive to
the exact value of the ratio.
The  line width in velocity space is proportional to the thermal velocity and is thus $\propto T^{1/2}$.
For the Paschen $\beta$ line on day 36
we find the observed FWHM to be {$\sim 2030 \kms$}, which is consistent with the following pairs of values of
($\tau_e$, $T$):  (1, 58600 K), (5, 16400 K), (10, 7760 K), and (20, 3700 K).

After interpolating over the narrow component of the day 36 Paschen $\beta$ line, we perform least squares fits of the electron scattering models to the broad component. We find the reduced $\chi^2$ to lie in the range $1.39 - 1.55$ for  $\tau_e$ ranging between 20 and 1 (Figures \ref{ESfits}). 
The data is best fit to the $\tau_e = 20$ profile, with a reduced $\chi^2$ of $1.39$. Such a high optical depth should be viewed with skepticism, especially in view of the small range of $\chi^2$ when $\tau_e$ varies between 1 and 20.

If the emitting and scattering gas are the same, another constraint on the emission comes from the ratio of the narrow line flux to the flux
in the broad wings, which is found to be 0.17 on day 36 for the Paschen $\beta$ line.
Using the model results from \cite{huang14},
we find that the ratio observed for SN 2010jl corresponds to $\tau_e \sim4$,
which suggests a lower limit of the scattering gas optical depth by assuming there is no other source of narrow line emission.
The observed temperature is then $\sim 20,000$ K, which is reasonable for the ionized gas.
Another measure of $\tau_e$ is provided by the outer wings of the line.
By adjusting the temperature, we can generate line profiles with the same FWHM at different $\tau_e$, the line wings become slightly stronger for higher values of $\tau_e$.
However, our data for SN 2010jl do not have sufficient signal-to-noise in the wings
to be able to use this method (Figure \ref{ESfits}),  as is indicated by the small range of 
reduced $\chi^2$ over the range of $\tau_e$ from 1 to 20.

Although a consistent picture for mixed emission and scattering can be developed for the
earliest time, it has problems at later times, indicating that application of the model to the narrow line at early times
may be misleading.
Over 100's of days, the broad component in the lines shifts to the blue, as also observed
in optical lines \citep{fransson13}.
One suggestion is that the shift is due to dust formation in the expanding medium \citep{smith12,maeda13,gall14}. 
In that picture, the redshifted photons emitted by material expanding away from the observer on the far side of the supernova are preferentially absorbed as they travel through a longer column of dust than their blueshifted near side counterparts. This would lead to a line profile that is less strong on the red side, an effect that would be more pronounced in the shorter wavelength lines as dust absorption within the ultraviolet/optical/near-infrared bands is more effective at shorter wavelengths. \cite{smith12} made this argument
based on spectra at an age $\sim 100$ days and \cite{maeda13} based on a spectrum at
age $563$ days.
At the earlier times, our spectra show a shift in the Brackett $\gamma$ line that is similar to that at
shorter wavelengths (Figure \ref{Hcompare}), although there is some uncertainty in the continuum level.
However, our observations cover a relatively small, long wavelength range and do not provide a
good test of wavelength dependent line profiles.

\cite{fransson13} discussed a number of problems with the shift in the broad component
being due to dust, and proposed a model in which there is radiative acceleration of the
gas  that gives rise to the broad component.
Since the narrow component does not shift to the blue along with the broad component,
the two components must originate in different places.
The equivalent width  in the narrow component of the H lines remains roughly constant over the first
200 days, so the ionized circumstellar medium appears to evolve slowly and the growth
in the equivalent width of the H lines is primarily due to the growth of the broad component.
Over the first 200 days, the NIR luminosity of SN 2010jl dropped by a factor of 3
\citep{fransson13}, so the narrow lines dropped by this factor and the flux in the
broad components approximately doubled from our earliest observation to an age of 200 days.
Formation of the narrow and broad lines  in the same place  would imply that the optical
depth increases with time  because fewer photons escape without scattering at higher optical depth.
This is unexpected, giving further evidence that the line
components form in separate regions.
The properties of the H emitting regions are distinct from those of the X-ray emitting region,
where the shock front has a velocity $\sim 4000\kms$ and a preshock
$\tau_e\sim 1$ on day 58 if there is incomplete ionization of the gas  \citep{chandra12,ofek14}.  The preshock column density subsequently
declines.
As discussed by \cite{fransson13}, the fact that the broad component shows a systematic
shift indicates that there is a large scale asymmetry in the object and that the optical
emission is perhaps from a polar flow.

As discussed in Section 2.3, the NIR He lines show a difference with the H lines.
In addition to the growth of a broad component shifted by $-1000\kms$, the He lines
develop a shoulder out to $-4000$ to $-5500\kms$.
The velocity indicates that this component may be associated with the higher
velocities inferred from the X-ray emission.
The fact that this feature does not appear in the H lines indicates a difference in
composition; the high velocity He feature may be associated with supernova ejecta
in which H is underabundant.

A surprisingly similar situation has been observed in the Type IIn SN 1997eg \citep{hoffman08}.
SN 1997eg had a maximum absolute $V$ magnitude  $\sim 1.5$ magnitudes fainter
than SN 2010jl, so a lower density interaction is indicated.
\cite{hoffman08} find that the H lines do not have Lorentzian profiles, implying that electron
scattering related to the thermal velocities of electrons is not important and consistent with a low density.
In addition, SN 1997eg was detected as a radio source at an age of 7 months \citep{lacey98}.
No detections of SN 2010jl have yet been reported, suggesting higher absorption in this case.
Despite the apparent difference
in circumstellar density, there are interesting similarities in the H and He lines.
In SN 1997eg, the H Balmer lines showed a shift in the peak of the line to the blue by
$\sim 700-800\kms$.
At the same time, the \ion{He}{1} $\lambda$7065 and $\lambda$5876 lines showed a strong shoulder
of emission to the blue by $\sim 5000\kms$.
At later times, a  peak like the one extending to the blue developed to the red in the H lines.
\cite{hoffman08} interpret the line emission as coming from an asymmetric region
with different geometries for the H and He dominated emission regions.
The H lines are attributed to shocks driven into a dense circumstellar disk or torus,
while the He emission is from H depleted gas that is expanding more rapidly into
a lower density circumstellar medium.
This view is supported by polarization observations of the supernova.

Our observations of H and He lines in SN 2010jl show similar shifts.
The H lines show only a blueshifted component, without the redshifted side.
We attribute the difference to the higher density and optical depths in the SN 2010jl case.
There is also evidence for blueshifted He rich gas moving at $5000\kms$.
However, the He emission is also strong in the lower velocity component, as opposed to
the case of SN 1997eg.

Our spectral line observations indicate
3 kinds of emitting regions:  a narrow line region of presupernova mass loss with velocities
$\sim 100\kms$, an 
intermediate line region involving $\sim 700\kms$ gas in an asymmetric structure, and    {\bf
a He dominant region of supernova ejecta with velocities $\sim 5000\kms$.
The lines are affected by electron scattering in the early phases of evolution and electron scattering may be responsible for the extended line wings through day $\sim 400$.
After day 400, the lines no longer show the characteristic electron scattering profiles and the line widths  probably reflect  gas velocities to $\sim 2000\kms$.
However,  the bulk of H rich material is moving at $\sim 700\kms$. 
The broad feature observed in the He lines could be produced by a uniformly
expanding shell of gas with velocity $\sim 5000\kms$; the absence of the redshifted part of the
shell may be due to the supernova being opaque.   }
Similar regions have been observed in other Type IIn supernovae, although the
high velocity ejecta region is sometimes observed to be O rich, e.g., SN 1986J \citep{mili08}
and SN 1995N \citep{fransson02}.

\section{DISCUSSION 	AND CONCLUSIONS}

Our study shows that NIR spectral observations of Type IIn supernovae can provide an
interesting window on these events.
For the H lines in SN 2010jl at early times, the NIR lines (Brackett and Paschen series) have
broad components with profiles that are similar to the broad components of the optical Balmer lines \citep{zhang12,fransson13}.
The line profiles are consistent with the lines being formed by electron scattering in the circumstellar medium.
The narrow line components in  the Balmer lines show absorption features, i.e. they are
optically thick, while the Paschen and Brackett lines are not.
The NIR narrow emission thus gives a measure of the flux of unscattered line photons emitted by
 the slow circumstellar medium.

In SN 2010jl, the optical \ion{He}{1} lines are relatively weak so there is not much information on
their profiles.
However, the NIR $\lambda$10830 line is strong and the profile is well defined in our observations.
We find a clear difference with the H lines in that there is a shoulder to the blue that implies
gas approaching at $4000-5500\kms$.
The emission can be interpreted as H poor supernova ejecta that are expanding into a region of
relatively small circumstellar density.
A similar situation is observed in SN 1997eg \citep{hoffman08}.

The NIR continuum can provide important information on the emission from warm dust, which appears
to be frequently present around Type IIn supernovae \citep{fox11}.
However, at early times the hotter photospheric emission is present and optical observations are
needed to separate out the photospheric emission from the dust emission \citep{fransson13}.
At later times, when the photospheric emission has faded, the NIR observations can provide 
crucial information on dust emission.

\acknowledgments
We are grateful to Ori Fox for help with the observations as well as discussion of the results.
We thank  Claes Fransson for  fruitful discussions and correspondence on SN 2010jl,
and the referee who provided detailed comments that led to significant improvement of the paper.
Thanks are also due to Meredith Drosback, Sarah Schmidt, and Yue Shen, who very graciously donated telescope time for our observations of SN 2010jl.
We also wish to thank Mike Skrutskie for his help with gathering, reducing, and interpreting the spectra.
This research was supported in part by NSF grant  AST-0807727 and NASA grant NNX12AF90G.
The research has made use of the SIMBAD database, operated at CDS, Strasbourg, France, and
 the NASA/IPAC Extragalactic Database (NED) which is operated by the Jet Propulsion Laboratory, California Institute of Technology, under contract with the National Aeronautics and Space Administration. 

{\it Facilities:} \facility{APO (TripleSpec)}.

\clearpage

\begin{deluxetable}{rccccr} 
\tablecolumns{6} 
\tablewidth{0pc} 
\tablecaption{Observing Details} 
\tablehead{ 
\colhead{Observation} & \colhead{Supernova}   & \colhead{Integration Time}    & \colhead{SN2010jl} & 
\colhead{HD85377}    & \colhead{Observers} \\
\colhead{Date} & \colhead{Age (days)} & \colhead{(minutes)} & \colhead{Airmass} & \colhead{Airmass} & }
\startdata 
2010 Nov 15  & 36 & 20 & 1.10 & 1.15 & M. Drosback, O. Fox \\
2010 Dec 02  & 53 & 20 & 1.10 & 1.16 & Y. Shen, O. Fox \\
2011 Jan 26  & 108 & 25 & 1.10 & 1.16 & G. Privon, O. Fox \\
2011 Feb 22  & 135 & 20 & 1.10 & 1.18 & Y. Shen, O. Fox \\
2011 Mar 23  & 164 & 20 & 1.32 & 1.47 & S. Schmidt, O. Fox \\
2011 Apr 06  & 178 & 20 & 1.13 & 1.17 & J. Borish, M. Skrutskie \\
2011 May 17  & 219 & 20 & 1.56 & 1.80 & J. Borish, O. Fox \\
2011 Nov 17  & 403 & 40 & 1.59 & 1.50 & J. Borish, B. Breslauer, \\
  &   &    &    &   &    A. Kingery \\
2012 Feb 09  & 488 & 40 & 1.14 & 1.24 & J. Borish \\
2012 Apr 26  & 565 & 40 & 2.0 & 2.1 & J. Borish \\
\enddata
\label{obstable}
\end{deluxetable}

\begin{deluxetable}{rcc} 
\tablecolumns{3} 
\tablewidth{0pc} 
\tablecaption{Blackbody Fits to the Near Infrared} 
\tablehead{ 
\colhead{Observation} & \colhead{Supernova}   & \colhead{Blackbody} \\
\colhead{Date} & \colhead{Age (days)} & \colhead{Temperature (K)} }
\startdata 
2011 Nov 17 & 403 & 2090  \\
2012 Feb 09 & 488 & 1890   \\
2012 Apr 26 & 565 &  1910  \\
\enddata
\tablecomments{Uncertainty in each temperature is approximately $\pm$75 K.}
\label{bbfits}
\end{deluxetable}

\clearpage

\begin{deluxetable}{rrrrrr} 
\tablecolumns{5}
\tablewidth{0pc} 
\tablecaption{Narrow to Broad Line Ratio} 
\tablehead{ 
\colhead{Date} & \colhead{Day}   & \colhead{Paschen $\beta$}    & \colhead{Paschen $\delta$} & 
\colhead{Brackett $\gamma$}}
\startdata 
2010 Nov 15 & 36 & 0.20$\pm$0.028 & 0.12$\pm$0.021 & 0.16$\pm$0.029 \\
2010 Dec 02 & 53 & 0.15$\pm$0.023 & 0.08$\pm$0.019 & 0.17$\pm$0.028 \\
2011 Jan 26 & 108 & 0.07$\pm$0.017 & 0.05$\pm$0.017 & 0.08$\pm$0.023 \\
2011 Feb 22 & 135 & 0.05$\pm$0.017 & 0.03$\pm$0.016 & 0.06$\pm$0.021 \\
2011 Mar 23 & 164 & 0.05$\pm$0.017 & 0.03$\pm$0.016 & 0.05$\pm$0.020 \\
2011 Apr 06 & 178 & 0.04$\pm$0.017 & 0.03$\pm$0.017 & 0.05$\pm$0.020 \\
2011 May 17 & 219 & 0.03$\pm$0.017 & 0.03$\pm$0.016 & 0.04$\pm$0.020 \\
2011 Nov 17 & 403 & 0.02$\pm$0.023 & 0.02$\pm$0.026 & 0.03$\pm$0.026 \\
2012 Feb 09 & 488 & 0.04$\pm$0.027 & 0.04$\pm$0.033 & 0.06$\pm$0.029 \\
2012 Apr 26 & 565 & 0.11$\pm$0.032 & \nodata & \nodata \\
\enddata
\label{n2b}
\end{deluxetable}

\begin{deluxetable}{cccccc} 
\tablecolumns{6} 
\tablewidth{0pc} 
\tablecaption{Equivalent Widths (Angstroms)} 
\tablehead{ 
\colhead{Date} & \colhead{Day}   & \colhead{Paschen $\beta$}    & \colhead{Paschen $\delta$} & 
\colhead{Brackett $\gamma$} }
\startdata 
2010 Nov 15 & 36 & 84$\pm$20 & 25$\pm$8 & 50$\pm$17 \\
2010 Dec 02 & 53 & 114$\pm$16 & 36$\pm$7 & 59$\pm$12 \\
2011 Jan 26 & 108 & 291$\pm$38 & 75$\pm$18 & 119$\pm$32 \\
2011 Feb 22 & 135 & 371$\pm$34 & 93$\pm$17 & 160$\pm$24 \\
2011 Mar 23 & 164 & 441$\pm$38 & 111$\pm$20 & 207$\pm$37 \\
2011 Apr 06 & 178 & 480$\pm$48 & 118$\pm$22 & 219$\pm$39 \\
2011 May 17 & 219 & 573$\pm$58 & 148$\pm$26 & 261$\pm$60 \\
2011 Nov 17 & 403 & 156$\pm$11 & 64$\pm$20 & 29$\pm$5 \\
2012 Feb 09 & 488 & 93$\pm$11 & 41$\pm$15 & 12$\pm$4 \\
2012 Apr 26 & 565 & 71$\pm$29 & \nodata & \nodata \\
\enddata
\tablecomments{ {\bf Broad and narrow line fluxes are included}}
\label{EWT}
\end{deluxetable}

\clearpage

\end{document}